\documentclass[twoside]{article}

\usepackage{anyfontsize}
\usepackage[T1]{fontenc} 
\usepackage[american]{babel}
\usepackage{RR}
\usepackage[toc,page]{appendix}
\usepackage{catchfilebetweentags}
\usepackage{cgitFigures}
\usepackage{BermudezRR2025}
\usepackage{catchfilebetweentags}

\RRNo{9591}
\RRdate{July 2025}
\RRauthor{
Yaiza Bermudez
\and Gaetan Bisson
\and Iñaki Esnaola
\and Samir~M.~Perlaza
}

\RRtitle{Démonstrations des théorèmes folkloriques sur la dérivée de Radon-Nikodym}
\RRetitle{Proofs for Folklore Theorems on the Radon-Nikodym Derivative}
\RRnote{
Yaiza Bermudez and Samir M. Perlaza  are with INRIA, Centre Inria d’Université Côte
d’Azur, Sophia Antipolis 06902, France. Gaetan Bisson and Samir M. Perlaza are with the
GAATI Laboratory at the Université de la Polynésie Française, Faaa 98702, French Polynesia.
Iñaki Esnaola is with the School of Electrical and Electronic Engineering at The University of Sheffield, Sheffield S1 3JD, UK.
Samir M. Perlaza and Iñaki Esnaola are also with the Electrical and Computer Engineering Department at Princeton University, Princeton N.J. 08544, USA. 
}

\RRresume{
	\foreignlanguage{french}{
Dans ce rapport technique, des énoncés rigoureux et des preuves formelles sont présentés pour des théorèmes folkloriques, tant fondamentaux qu’avancés, portant sur la dérivée de Radon-Nikodym. Les cas des mesures de probabilité conditionnelles et marginales sont examinés avec soin, ce qui conduit à une identité impliquant la somme de l’information mutuelle et de l’information lautum, suggérant une nouvelle interprétation de celle-ci.
	}
 }
 
\RRabstract{ 
  In this technical report, rigorous statements and formal proofs are presented for both foundational and advanced folklore theorems on the Radon-Nikodym derivative. The cases of conditional and marginal probability measures are carefully considered, which leads to an identity involving the sum of mutual and lautum information suggesting a new interpretation for such a sum.
}
 
\RRmotcle{\foreignlanguage{french}{Dérivé de Radon Nikodym}}
\RRkeyword{Radon Nikodym Derivative.}

\RRprojet{NEO}  
\RCSophia 

\usepackage{pdfpages} 
\usepackage[section]{placeins}
\usepackage{float}
\usepackage{pdfpages} 
\allowdisplaybreaks

\title{Folklore Theorems of the Radon-Nikodym Derivative} 	
\author{Yaiza Bermudez
\and Gaetan Bisson
\and Samir~M.~Perlaza
\and I{\~n}aki Esnaola
} 		
\date{\today} 				
\begin{document}
\makeRR   
%
\clearpage
\tableofcontents
\clearpage
\selectlanguage{american}
%
\section{Introduction}
\IEEEPARstart{I}{n} mathematics, folklore theorems refer to results that are widely accepted and frequently utilized by experts but are often not  formally proven or explicitly documented. In game theory for example, the original Folk theorem earned its name because, although it was widely recognized among game theorists during the 1950s, it remains unpublished and without attribution to particular authors. See for instance~\cite{TheFolkTheorem} and~\cite{fudenberg1991game}.
Folklore theorems populate all areas in mathematics. 
In information theory, a large set of folklore theorems involve the Radon-Nikodym derivative (RND),  first introduced by Radon~\cite{radon1913theorie}; and later generalized by Nikodym~\cite{nikodym1930generalisation}. 
%
%
The existence of such folklore theorems in this area arises in part from the fact that all Shannon's information measures can be defined in terms of the RND. Interestingly, most properties of the RND are often presented without proof in most textbooks on measure theory and probability theory, c.f, \cite{ash2000probability, halmos1950measure, billingsley2012probability, royden2010real, axler2020measure, durrett2019probability, lehmann2005testing}, with the exception of some properties presented in~\cite{FollandRealAnalysis}.

Claude Shannon did not use the RND in his foundational papers \cite{Shannon-1948a, Shannon-1948b} to define entropy and mutual information. Instead, Shannon opted for restricting his work to the case in which measures either possess a probability mass function or a probability density function, which are both instances of RNDs. 
This choice significantly influenced the presentation of most subsequent results in information theory and established the style in which classical textbooks were written \cite{kullback1959information, cover2006elements, ash2012information, bloch2011physical, cinlar2011probability, csiszar2011information, elgamal2011network, fano1961transmission, feinstein1958foundations,  gallager1968information, han2003information, kapur1989maximum, mceliece2002theory, mezard2009information, pinsker1964information, wolfowitz1964coding, yeung2008information}. 
 Nonetheless, the RND has been increasingly adopted in modern textbooks \cite{polyanskiy2024information} and in the definition of new information measures, e.g., lautum information \cite{palomar2008lautum}, to privilege a unified presentation. That is, independently of the measure used as a reference, e.g., the counting measure, the Lesbegue measure, etc. Some recent results whose presentation relies on the RND are for instance \cite{sason2016fdivergence, daunas2024equivalence, daunas2024asymmetry, agrawal2021optimal, asadi2020chaining, masiha2023fdivergence, rodriguezgalvez2024information, yavas2023variable, perlaza2024generalization, verdu2024relative, zouJSAIT2024, PerlazaTIT2024}.

At the light of this observation, this technical report presents rigorous statements and alternative proofs for the most common folklore theorems on the RND. These include the change of measure theorem, the proportional-measures theorem, the chain rule, the multiplicative inverse, linearity and continuity theorems, as well as the product-measure theorem and nonnegativity and finiteness theorem. Less common theorems such as the unit measure and two Bayes-like rules of the RND are also formally proved. 
Finally, these novel theorems are used to prove a key identity involving the sum of mutual and lautum information, which is often observed in statistical learning \cite{aminian2022information, perlaza2024generalization}.
Hopefully, these contributions would provide a valuable reference for researchers and students in information theory.

\section{Preliminaries}
This section introduces relevant notational conventions alongside the Radon-Nikodym theorem.
In particular, some equalities presented in this paper are valid almost surely with respect to a given measure. For clarity, given a measure space $\left( \Omega, \mathscr{F}, P \right)$, the notation $\eqasP$ is introduced and shall be read as ``equal for all $x \in \Omega$ except on a negligible set with respect to $P$''; or equivalently as ``equal almost surely with respect to $P$''. Moreover, given two measures $P$ and $Q$ on the same measurable space, the notation $\abscontPQ$ stands for ``the measure $P$ is absolutely continuous with respect to $Q$''.
Using this notation, the Radon-Nikodym derivative is introduced by the following theorem.
\begin{theorem}[Radon-Nikodym theorem, {\cite[Theorem~2.2.1]{ash2000probability}}]\label{ThRNT}
Let~$P$ be a signed measure and~$Q$ be a $\sigma$-finite measure both on a given measurable space~$\left( \Omega, \mathscr{F}\right)$, with 
$ \abscontPQ$. Then, there exists a Borel measurable 
function~$g:~\Omega~\to~\mathbb{R}$ such that for all~$\mathcal{A}\in\mathscr{F}$, 
\begin{equation}\label{EqRNDa}
 P(\mathcal{A}) = \int_{\mathcal{A}} g(x) \mathrm{d}Q(x).
\end{equation}
Moreover, if another function $h$ satisfies for all~$\mathcal{A} \in \mathscr{F}$ that~$P(\mathcal{A}) = \int_{\mathcal{A}} h(x) 
\mathrm{d}Q(x)$, then $ g(x) \eqas{Q}$$ h(x)$. 
\end{theorem}
The function $g$ in \eqref{EqRNDa} is often referred to as the Radom-Nikodym derivative of $P$ with respect to $Q$; and is also written as~$\frac{\mathrm{d}P}
{\mathrm{d}Q} $, such that~$g(x)~=~\frac{\mathrm{d}P}{\mathrm{d}Q} (x).$

The Radon-Nikodym derivative exhibits a number of properties from which many foundational results originate. Some of these properties are certainly folklore theorems and thus, the purpose of the following section is to provide formal proofs. Other properties are formally proved in literature and thus,  for the sake of completeness, an alternative proof is presented using the described notation.

\section{Basic Folklore Theorems}
This section focuses on basic folklore theorems, where ``basic'' denotes their well-established nature. One of the most common folklore theorems is often referred to as the ``change of measure'' theorem.
%
\begin{theorem}[Change of Measure]\label{ThCOM}
Let~$P$ be a signed measure and~$Q$ be a $\sigma$-finite measure both on a given measurable space~$\left( \Omega, \mathscr{F}\right)$, with $\abscontPQ$. Let $f: \Omega \to \reals$ be a Borel 
measurable function such that the integral $\int_{\Omega} f(x) \mathrm{d}P(x)$ exists. Then,  for all $
\mathcal{A} \in \mathscr{F}$,
\begin{equation}\label{EqTheoRNProperties1}
\displaystyle\int_{\mathcal{A}} f(x)\mathrm{d}P(x) = \displaystyle\int_{\mathcal{A}} f(x) \frac{\mathrm{d}P}{\mathrm{d}Q}(x)
 \mathrm{d}Q(x).
\end{equation} 
\end{theorem}
A proof of this theorem is presented in \cite[Proposition~3.9]{FollandRealAnalysis}. For the sake of completeness, an alternative proof is presented hereunder, using the previously described notation.
\begin{IEEEproof}
Assume that $f$ is a simple function, i.e.\ for all $x \in \Omega$,
$f(x) = \sum_{i=1}^{m} a_i \mathds{1}_{\mathcal{A}_i}(x),$ for some finite $m \in \mathbb{N}$, disjoint sets $\mathcal{A}_1,\ldots,\mathcal{A}_m$ in $\mathscr{F}$, and reals $a_1,\ldots,a_m$. For all $\mathcal{A}\in\mathscr{F}$ and $i\in\{1,\ldots,m\}$, let $\mathcal{B}_i = \mathcal{A}\cap \mathcal{A}_i$. Then,
\begin{IEEEeqnarray}{rCl}
\label{EqProofPart1.1ChangeMeasure}
&& \int_{\mathcal{A}} f(x)  \frac{\mathrm{d}P}{\mathrm{d}Q}(x) \mathrm{d} Q (x)
 =  \int_{\mathcal{A}} \frac{\mathrm{d}P}{\mathrm{d}Q} (x) \sum_{i=1}^{n} a_i \mathds{1}_{\mathcal{A}_i} (x) \mathrm{d} Q(x)\\
\label{EqProofPart1.2ChangeMeasure}
&  = & \sum_{i=1}^{n} a_i \int_{\mathcal{B}_i}  \frac{\mathrm{d}P}{\mathrm{d}Q}(x) \mathrm{d} Q(x) 
  =  \sum_{i=1}^{n} a_i P (\mathcal{B}_i),
\end{IEEEeqnarray} 
where the equality in \eqref{EqProofPart1.2ChangeMeasure} follows Theorem~\ref{ThRNT} and the linearity of the integral \cite[Theorem~1.6.3]{ash2000probability}. On the other hand, for all $\set{A} \in \mathscr{F}$,
\begin{IEEEeqnarray}{rCl}
\label{EqProofPart2ChangeMeasure}
\int_{\mathcal{A}} f(x) \,\mathrm{d}P(x) 
&=& 
\sum_{i=1}^{m} a_i \int_{\mathcal{B}_i} \mathrm{d}P(x) 
\,=\,
\sum_{i=1}^{m} a_i P(\mathcal{B}_i),
\end{IEEEeqnarray}
where the equality in \eqref{EqProofPart2ChangeMeasure} follows from the linearity of the integral \cite[Theorem~1.6.3] {ash2000probability}. Hence, from Theorem~\ref{ThRNT} when $f$ is simple, the desired equality in~\eqref{EqTheoRNProperties1} holds immediately.

The proof proceeds by observing that: \emph{(a)} simple functions are dense in the space of Borel measurable functions \cite[Theorem~1.5.5(b)]{ash2000probability}, and \emph{(b)} the integral is a continuous map on that space \cite[Theorem~1.6.2]{ash2000probability}.  From \emph{(a)} and \emph{(b)}, it follows that \eqref{EqTheoRNProperties1} also holds for any Borel measurable function~$f$. This completes the proof.
\end{IEEEproof}
Another reputed  folklore theorem, which  is often referred to as the ``proportional measures’’ theorem, establishes the explicit form of the Radon-Nikodym derivative between two measures, in which one is proportional to the other.
\begin{theorem}[Proportional Measures]\label{TheoRNDOne}
Let $P$ and $Q$ be two $\sigma$-
finite measures on the measurable space $\left( \Omega, \mathscr{F}\right)$, such that for all $\set{A} \in \mathscr{F}$
\begin{equation}
\label{EqEqualMeasuresUpToConstant}
Q(\set{A}) = c P (\set{A}),
\end{equation}
with $c > 0$. Then,
\begin{equation}
\label{EqThPropconst}
\frac{\mathrm{d}P}{\mathrm{d}Q} (x) \eqas{Q} \frac{1}{c},\quad\text{and}\quad\frac{\mathrm{d}Q}{\mathrm{d}P} (x) \eqas{P}  c.
\end{equation}
\end{theorem}
\begin{IEEEproof}
Note that from \eqref{EqEqualMeasuresUpToConstant}, $P$ and $Q$ are mutually absolutely continuous. Moreover, from Theorem~\ref{ThRNT}, for all $\mathcal{A}\in\mathscr{F}$,
\begin{equation}
P(\mathcal{A}) = \int_{\mathcal{A}} \mathrm{d}P(x) \;=\; \int_{\mathcal{A}} \frac{\mathrm{d}P}{\mathrm{d}Q}(x)\,\mathrm{d}Q(x),
\end{equation}
and from \eqref{EqEqualMeasuresUpToConstant}, it follows that
\begin{equation}
P(\mathcal{A}) = \frac{1}{c}\,Q(\mathcal{A}) \;=\; \int_{\mathcal{A}} \frac{1}{c}\,\mathrm{d}Q(x).
\end{equation}
Hence, $\frac{\mathrm{d}P}{\mathrm{d}Q}(x) \eqas{Q} \frac{1}{c}$ follows directly from Theorem~\ref{ThRNT}. A similar argument shows that $\frac{\mathrm{d}Q}{\mathrm{d}P}(x) \eqas{P} c$.
\end{IEEEproof}
\noindent
In the special case $c=1$ in \eqref{EqThPropconst}, $P$ and $Q$ are identical, yielding $ \frac{\mathrm{d} P}{\mathrm{d} Q}(x) \eqas{Q}$ $ \frac{\mathrm{d} Q}{\mathrm{d} P}(x) \eqasP $ $1.$

The following folklore theorem is often referred to as the ``chain rule'' and a proof appears in \cite[Proposition~3.9]{FollandRealAnalysis}. For the sake of completeness, an alternative proof is presented hereunder, using the previously described notation.
 \begin{theorem}[Chain Rule]\label{TheoChainsAndBlood}
Let $P$ be a signed measure and let $Q$ and $R$ be two $\sigma$-finite measures, all  on the same measurable space $\left( \Omega, \mathscr{F}\right)$, with
$\abscontPQ$ and
$\abscont{Q}{R}$.
Then, 
\begin{equation}\label{EqLimitedEditions1}
\frac{\mathrm{d}P}{\mathrm{d}R}(x) \eqas{R} \frac{\mathrm{d}P}{\mathrm{d}Q}(x)  \frac{\mathrm{d}Q}{\mathrm{d}R}(x).
\end{equation}
\end{theorem}
\begin{IEEEproof}
From the assumptions of the theorem and Theorem~\ref{ThCOM}, it follows that for all $\mathcal{A}\in \mathscr{F}$,
\begin{IEEEeqnarray}{rCl}
P(\mathcal{A})
&=&
\int_{\mathcal{A}} \mathrm{d}P(x)
\,=\,
\int_{\mathcal{A}} \frac{\mathrm{d}P}{\mathrm{d}Q}(x)\,\mathrm{d}Q(x)\\
\label{EqChainRule3}
&=&
\int_{\mathcal{A}} \frac{\mathrm{d}P}{\mathrm{d}Q}(x)\,\frac{\mathrm{d}Q}{\mathrm{d}R}(x)\,\mathrm{d}R(x)
\,=\,
\int_{\mathcal{A}} \frac{\mathrm{d}P}{\mathrm{d}R}(x)\,\mathrm{d}R(x),
\end{IEEEeqnarray}
where the second equality in \eqref{EqChainRule3} holds given that $P\ll R$ and Theorem~\ref{ThRNT} which yields \eqref{EqLimitedEditions1}.
\end{IEEEproof}
The following folklore theorem is a consequence of Theorem~\ref{TheoChainsAndBlood} and shows the connection between the Radon-Nikodym derivative and its multiplicative inverse.%
\begin{theorem}[Multiplicative Inverse]\label{TheoInverseRND}
Let $P$ and $Q$ be two mutually absolutely continuous $\sigma$-finite measures on the measurable space $\left( \Omega, \mathscr{F}\right)$; and assume that for all $x \in \Omega$,  $\frac{\mathrm{d}Q}{\mathrm{d}P}(x) > 0$. 
Then,  
\begin{IEEEeqnarray}{rCl}\label{EqLimitedEditions}
\frac{\mathrm{d}P}{\mathrm{d}Q}(x) &  \eqas{Q} & \left( \frac{\mathrm{d}Q}{\mathrm{d}P}(x) \right)^{-1}.
\end{IEEEeqnarray}
\end{theorem}
This result also appears in \cite[Corollary~3.10]{FollandRealAnalysis}. 
\begin{IEEEproof}
From Theorem~\ref{TheoChainsAndBlood}, it follows that 
\begin{IEEEeqnarray}{rCl}
\label{EqSomebodyCalls}
\frac{\mathrm{d} P}{\mathrm{d} Q} (x) \frac{\mathrm{d} Q}{\mathrm{d} P} (x)  \eqas{Q} \frac{\mathrm{d} Q}{\mathrm{d}Q}(x) \eqas{Q}  1,
\end{IEEEeqnarray}
where the last equality follows from Theorem~\ref{TheoRNDOne}, with~$c~=~1$.
\end{IEEEproof}
The subsequent folklore theorem establishes the linearity of the Radon-Nikodym derivative. A similar statement is presented in \cite[Proposition~3.11]{FollandRealAnalysis} but left without proof. 
\begin{theorem}[Linearity]\label{TheoRNDadditivity}
Let~$P$ be a $\sigma$-finite measure on~$\left(\Omega, \mathscr{F}\right)$ and let also~$Q_1 , Q_2 , \ldots , Q_n$ be signed measures on~$\left(\Omega, \mathscr{F}\right)$ absolutely continuous with respect to~$P$. Let~$c_1 , c_2 , \ldots , c_n$ be positive reals; and let $S$ be a signed measure on $\left(\Omega, \mathscr{F}\right)$ such that for all~$\set{A} \in \mathscr{F}$, 
$S(\set{A}) = \sum_{t=1}^n c_t Q_t(\set{A})$.
Then,  
\begin{equation}
\label{EqJanuary18at22h41inDubai}
\frac{\mathrm{d} S}{\mathrm{d}P}(x)  \eqasP  \sum_{t =1}^{n} c_t \frac{\mathrm{d}Q_t}{\mathrm{d}P} (x).
\end{equation}
\end{theorem}
\begin{IEEEproof}
From the assumptions of the theorem, it holds that $\abscont{S}{P}$, thus for all~$\set{A} \in \mathscr{F}$,
\begin{IEEEeqnarray}{rCl} \label{EqProofTheoRNDadditivity2}
\label{EqChangeofmeasure105}
 &\int_{\mathcal{A}}& \frac{\mathrm{d} S}{\mathrm{d}P}(x) \mathrm{d} P(x) = \int_{\mathcal{A}}\mathrm{d} S(x) = \sum_{t=1}^n c_t \int_{\mathcal{A}} \mathrm{d} Q_t(x) \spnum\\ 
\label{EqChangeMeasure}
&= & \sum_{t=1}^n \int_{\mathcal{A}} c_t \frac{ \mathrm{d} Q_t}{\mathrm{d}P}(x) \mathrm{d} P(x) = \int_{\mathcal{A}} \sum_{t=1}^n c_t \frac{ \mathrm{d} Q_t}{\mathrm{d}P}(x) \mathrm{d} P(x),
\end{IEEEeqnarray}
where the first equality in~\eqref{EqChangeofmeasure105} and  the first equality in~\eqref{EqChangeMeasure} follow from Theorem~\ref{ThCOM}; and the last equality in 
\eqref{EqChangeMeasure} follows from the additivity property of the integral \cite[Corollary~1.6.4]{ash2000probability}.
The proof ends by using Theorem~\ref{ThRNT} yielding~\eqref{EqJanuary18at22h41inDubai}.
\end{IEEEproof}
The following folklore theorem establishes the continuity of the Radon-Nikodym derivative.
\begin{theorem}[Continuity]
Let~$P$ be a $\sigma$-finite measure on $\left(\Omega, \mathscr{F} \right)$, and let $Q_1, Q_2, \cdots$ be an infinite
sequence of finite measures on $\left( \Omega, \mathscr{F} \right)$, converging to a signed measure $Q$. Suppose that for all $n \in \mathbb{N}$, $\abscont{Q_n}{P}$.
Then, $\abscont{Q}{P}$ and 
\begin{equation}\label{EqTheoRNDadditivity1}
\displaystyle\lim_{n \rightarrow \infty}  \frac{ \mathrm{d} Q_n}{\mathrm{d}P}(x) \eqasP  \frac{\mathrm{d}Q}{\mathrm{d}P}(x).
\end{equation}
\end{theorem}
\begin{IEEEproof}
From the assumptions of the theorem and Theorem~\ref{ThCOM}, for all  $\mathcal{A} \in \mathscr{F}$, it holds that
\begin{IEEEeqnarray}{rCl}
\label{EqProofThmContinuity}
Q(\mathcal{A}) &=& \lim_{n\to\infty} Q_n(\mathcal{A})
\;=\;
\int_{\mathcal{A}} \lim_{n\to\infty} \frac{\mathrm{d}Q_n}{\mathrm{d}P}(x)\,\mathrm{d}P(x),
\end{IEEEeqnarray}
where the last equality in \eqref{EqProofThmContinuity}  follows from Theorem~\ref{ThCOM} and \cite[Theorem~1.6.2]{ash2000probability}. Now, from the assumption that $Q\ll P$, it follows that for all  $\mathcal{A} \in \mathscr{F}$, 
\begin{equation}
\label{EqProofThmContinuity2}
Q(\mathcal{A}) \;=\; \int_{\mathcal{A}} \frac{\mathrm{d}Q}{\mathrm{d}P}(x)\,\mathrm{d}P(x).
\end{equation}
Therefore, the equality in \eqref{EqProofThmContinuity2} and Theorem~\ref{ThRNT} yields~\eqref{EqTheoRNDadditivity1}.
\end{IEEEproof}
The ensuing folklore theorem relates the Radon–Nikodym derivative of a product measure to those of its component measures. This result also appears in \cite[Exercise~3.12]{FollandRealAnalysis}.
 \begin{theorem}[Product of Measures]
For all $i \in \lbrace 1,2 \rbrace$, let $P_{i}$ and $Q_{i}$ be a signed and a $\sigma$-finite measure on $\left( \Omega_i, \mathscr{F}_i \right)$, respectively; with $\abscont{P_i}{Q_i}$.  Let also $P_1P_2$ and 
$Q_1 Q_2$ be the product measures on $\left( \Omega_1 \times \Omega_2, \mathscr{F}_1 \times \mathscr{F}_2 \right)$ formed by $P_1$ and $P_2$; and $Q_1$ and $Q_2$, 
respectively. Then, 
 \begin{IEEEeqnarray}{rCl}
  \label{EqProductMeasure0}
\frac{\mathrm{d}P_{1} P_{2}}{\mathrm{d}Q_1Q_2}\left(x_1, x_2\right) &\eqas{Q_1Q_2}& \frac{\mathrm{d}P_{1}}{\mathrm{d}Q_1} \left(x_1\right)\frac{\mathrm{d}P_{2}}{\mathrm{d}Q_2}\left(x_2\right).
\end{IEEEeqnarray}
 \end{theorem}
 \begin{IEEEproof}
By definition of product measures, for all $\mathcal{A}\subseteq \Omega_1\times\Omega_2$,
\begin{IEEEeqnarray}{rCl}
\label{EqProductMeasure1}
P_{1} P_{2}\left(\set{A}\right) &=& \int_{\set{A}} \mathrm{d}P_{1} P_{2}\left(x_1, x_2\right) 
= \iint_{\set{A}_{x_2}} \mathrm{d}P_{1}\left(x_1\right) \mathrm{d}P_{2}\left(x_2\right)\spnum\\
 \label{EqProductMeasure3}
&=& \iint_{\set{A}_{x_2}}\dfrac{\mathrm{d}P_{1}\left(x_1\right)}{\mathrm{d}Q_1} \mathrm{d}Q_1\left(x_1\right) \mathrm{d}P_{2}\left(x_2\right)\\
 \label{EqProductMeasure4}
&=& \iint_{\set{A}_{x_2}} \dfrac{\mathrm{d}P_{1}\left(x_1\right)}{\mathrm{d}Q_1} \dfrac{\mathrm{d}P_{2}\left(x_2\right)}{\mathrm{d}Q_2} \mathrm{d}Q_1\left(x_1\right) \mathrm{d}Q_2\left(x_2\right) \IEEEeqnarraynumspace\\
 \label{EqProductMeasure5}
&=& \int_{\set{A}} \dfrac{\mathrm{d}P_{1}}{\mathrm{d}Q_1}\left(x_1\right) \dfrac{\mathrm{d}P_{2}}{\mathrm{d}Q_2}\left(x_2\right) \mathrm{d}Q_1Q_2\left(x_1, x_2\right), 
\end{IEEEeqnarray}
where~$\set{A}_{x_2}$ is the section of the set~$\set{A}$ determined by~$x_2$, namely, $\mathcal{A}_{x_2}=\{x_1 : (x_1,x_2)\in\mathcal{A}\}$; the equality in \eqref{EqProductMeasure1} arises from the definition of $P_1P_2$  as the product of $P_1 $ and  $P_2$;  the equalities in \eqref{EqProductMeasure3} and \eqref{EqProductMeasure4} are direct consequences of Theorem~\ref{ThCOM}; and finally, the equality in \eqref{EqProductMeasure5} is due to the construction of  $Q_1Q_2$ as the product measure of  $Q_1$ and $Q_2$.
From the equality in \eqref{EqProductMeasure5}, it holds that $\abscont{P_1P_2}{Q_1Q_2}$. Thus, for all~$\set{A}~\in~\mathscr{F}_1~\times~\mathscr{F}_2$, 
\begin{equation}
\label{EqCartProduct1}
P_{1} P_{2}\left(\set{A}\right) = \int_{\set{A}} \dfrac{\mathrm{d}P_{1} P_{2}}{\mathrm{d}Q_1Q_2}\left(x_1, x_2\right) \mathrm{d}Q_1Q_2\left(x_1, x_2\right).
\end{equation}

From Theorem~\ref{ThRNT}, the equality in \eqref{EqProductMeasure0} follows, completing the proof.
\end{IEEEproof}
The following theorem shows a nonnegativity and finiteness property of the RND. 
\begin{theorem}[Nonnegativity and Finiteness]
Consider a probability measure $P$ and a $\sigma$-finite measure $Q$  on $\left( \Omega, \mathscr{F} \right)$, with $P\ll Q$. Then 
\begin{equation}
P\autoparent{\left\lbrace \omega \in \Omega :  0 \leqslant \RND{P}{Q}(\omega)  < +\infty \right\rbrace} = 1.
\end{equation}
\end{theorem}
\begin{IEEEproof}
Assume that 
\begin{equation}
P\autoparent{\left\lbrace \omega \in \Omega :  \RND{P}{Q}(\omega)  < 0 \right\rbrace} > 0.
\end{equation}
Hence, 
\begin{IEEEeqnarray}{rCl}
P\autoparent{\left\lbrace \omega \in \Omega :  \RND{P}{Q}(\omega)  < 0 \right\rbrace} & = & \int_{\left\lbrace \omega \in \Omega :  \RND{P}{Q}(\omega)  < 0 \right\rbrace} \mathrm{d}P(x)\\
& = & \int_{\left\lbrace \omega \in \Omega :  \RND{P}{Q}(\omega)  < 0 \right\rbrace}  \RND{P}{Q}(x)  \mathrm{d}Q(x)\\
& < & 0,
\end{IEEEeqnarray}
which contradicts that $P$ is a probability measure, which is nonnegative. 
Alternative, assume that 
\begin{equation}
P\autoparent{\left\lbrace \omega \in \Omega :  \RND{P}{Q}(\omega)  = +\infty \right\rbrace} > 0.
\end{equation}
Hence, 
\begin{IEEEeqnarray}{rCl}
P\autoparent{\left\lbrace \omega \in \Omega :  \RND{P}{Q}(\omega)  = +\infty \right\rbrace} & = & \int_{\left\lbrace \omega \in \Omega :  \RND{P}{Q}(\omega)  = +\infty \right\rbrace} \mathrm{d}P(x)\\
& = & \int_{\left\lbrace \omega \in \Omega :  \RND{P}{Q}(\omega)  = +\infty \right\rbrace}  \RND{P}{Q}(x)  \mathrm{d}Q(x)\\
& = & +\infty,
\end{IEEEeqnarray}
which contradicts that $P$ is a probability measure, which is a finite measure.
\end{IEEEproof}
\section{Advanced Folklore Theorems}
This section focuses exclusively on probability measures with the aim of establishing closer ties to information measures, which requires some additional notation.
 In particular, denote by $\triangle\bigl(\mathcal{X},\mathscr{F}_\mathcal{X}\bigr)$, or simply~$\triangle\bigl(\mathcal{X}\bigr)$, the set of all probability measures on the measurable space $\bigl(\mathcal{X},\mathscr{F}_{\set{X}}\bigr)$, where $\mathscr{F}_{\set{X}}$ is a $\sigma$-algebra on $\set{X}$.
Using this notation, conditional probability measures can be defined as follows.
\begin{definition}[Conditional Probability]
A family $P_{Y|X}\,\triangleq\,(P_{Y|X=x})_{x\in\set{X}}$ of elements of $\triangle\bigl(\set{Y}, \mathscr{F}_{\set{Y}}\bigr)$ indexed by $\set{X}$ is said to be a conditional probability measure if, for all sets $\mathcal{B} \in \mathscr{F}_{\set{Y}}$, the map
\begin{equation}
\function{\set{X}}{[0,1]}{x}{P_{Y|X=x}(\mathcal{B})}
\end{equation}
is Borel measurable. The set of such conditional probability measures is denoted by $\triangle(\set{Y}\!\mid\!\set{X})$.
\end{definition}
A conditional probability $P_{Y|X} \in \triangle(\mathcal{Y} \mid \mathcal{X})$ and a probability measure $P_X \in \triangle(\mathcal{X})$ jointly determine two unique probability measures in $\triangle(\mathcal{X}\times\mathcal{Y})$ and $\triangle(\mathcal{Y}\times\mathcal{X})$, respectively. These probability measures are denoted by $P_{XY}$ and $P_{YX}$, respectively, and for all sets $\mathcal{A} \in \mathscr{F}_{\set{X}} \times \mathscr{F}_{\set{Y}}$, it follows that
\begin{IEEEeqnarray}{rCl}
\label{EqMay20at14h23in2024}
P_{X Y }\bigl(\set{A}\bigr)
& = &
\int P_{Y |X = x} \bigl(\set{A}_{x}\bigr)\,\mathrm{d} P_{X} (x),
\end{IEEEeqnarray}
where~$\set{A}_{x}$ is the section of the set~$\set{A}$ determined by~$x$, namely,
\begin{IEEEeqnarray}{rCl}
\label{EqNovember15at16h57in2024InTheBusToNice}
\set{A}_{x}
& \triangleq &
\bigl\lbrace y \in \set{Y}: (x,y) \in \set{A} \bigr\rbrace.
\end{IEEEeqnarray}
Alternatively, for all sets~$\set{B} \in \mathscr{F}_{\set{Y}} \times \mathscr{F}_{\set{X}}$, it follows that
\begin{equation}
\label{EqOctober30at7h48in2024SophiaAntipolis}
P_{ YX } \bigl( \set{B}\bigr)= \int P_{Y |X = x} \bigl(\set{B}_{x}\bigr)\,\mathrm{d} P_{X} (x),
\end{equation}
where~$\set{B}_{x}$ is the section of the set~$\set{B}$ determined by~$x$.
\noindent
Note that \(P_{XY}\) is a measure in $\simplex{\set{X} \times \set{Y}}$ whereas \(P_{YX}\) is a measure in $\simplex{\set{Y} \times \set{X}}$.
Consequently, obtaining \(P_{XY}\) from \(P_{YX}\) requires more than simply renaming \(X\) and \(Y\); one must also properly redefine the corresponding measurable spaces.
In particular, given a set~$\set{A} \in  \mathscr{F}_{\set{X}} \times \mathscr{F}_{\set{Y}}$, let the set~$\hat{\set{A}} \in  \mathscr{F}_{\set{Y}} \times \mathscr{F}_{\set{X}}$ be such that 
\begin{equation}
\label{EqNovember16at18h49in2024Nice}
\hat{\set{A}} = \bigl\lbrace (y,x) \in \set{Y}\times\set{X} : (x,y)\in \set{A}\bigr\rbrace.
\end{equation}
Then, from~\eqref{EqMay20at14h23in2024} and~\eqref{EqOctober30at7h48in2024SophiaAntipolis}, it holds that
\begin{equation}
\label{EqJanuary10at16h43in2025}
P_{XY}\bigl(\set{A}\bigr) = P_{YX}\bigl(\hat{\set{A}}\bigr).
\end{equation}
Using this notation, a marginal probability measure is defined as follows.
\begin{definition}[Marginal]
Given two joint probability measures $P_{XY} \in \triangle\left( \set{X} \times \set{Y} \right)$ and $P_{YX} \in \triangle\left( \set{Y} \times \set{X} \right)$, satisfying \eqref{EqJanuary10at16h43in2025}, the marginal probability measures in $\triangle\left( \set{X} \right)$ 
and $\triangle\left( \set{Y} \right)$, 
denoted by $P_X$ 
and $P_Y$, respectively, 
satisfy for all  sets $\set{A} \in  \mathscr{F}_{\set{X}}$ 
\begin{equation}
\label{eqmargpx}
P_{X}\left( \set{A} \right)  \triangleq  P_{XY}\left( \set{A} \times \set{Y} \right) =  P_{YX}\left( \set{Y} \times \set{A} \right);
\end{equation}
and for all sets $\set{B} \in \mathscr{F}_{\set{Y}}$,
\begin{equation}
\label{eqmargpy}
P_{Y}\left( \set{B} \right)  \triangleq  P_{YX}\left( \set{B} \times \set{X} \right) =  P_{XY}\left( \set{X} \times \set{B} \right).
\end{equation}
\end{definition}

From the total probability theorem \cite[Theorem~4.5.2]{ash2000probability}, it follows that for all~$\set{B} \in \mathscr{F}
_{\set{Y}}$,  
\begin{equation}
\label{eqmarginal1}
P_Y(\set{B})  = \iint_{\set{B}} \mathrm{d}P_{Y | X = x}(y) \mathrm{d}P_X (x);
\end{equation}
and for all $\set{A} \in \mathscr{F}_{\set{X}}$,
\begin{equation}
\label{eqmarginal2}
P_X(\set{A}) = \iint_{\set{A}} \mathrm{d}P_{X | Y = y}(x) \mathrm{d}P_Y(y).
\end{equation}
In a nutshell, the joint probability measures~$P_{XY}$ and~$P_{YX}$ can be described via the conditional probability measure~$P_{Y | X}$ 
and the probability measure~$P_{X}$ as in~\eqref{EqMay20at14h23in2024} and 
in~\eqref{EqOctober30at7h48in2024SophiaAntipolis}; or via the conditional probability measure~$P_{X | Y} \in \triangle\left( \set{X} | \set{Y} \right)$ and the marginal probability 
measure~$P_{Y} \in \triangle\left( \set{Y} \right)$.
More specifically, for all sets~$\set{A} \in  \mathscr{F}_{\set{X}} \times \mathscr{F}_{\set{Y}}$, it follows that
\begin{equation}
\label{EqJun3at14h31in2024}
P_{XY }\left( \set{A} \right) = \int P_{X | Y = y} \left( \set{A}_{y}\right) \mathrm{d} P_{Y} \left( y\right),
\end{equation}
where~$\set{A}_{y}$ is the section of the set~$\set{A}$ determined by~$y$, namely, 
\begin{equation}
\label{EqNovember15at16h39in2024InTheBusToNice}
\set{A}_{y} \triangleq \left\lbrace x \in \set{X} : (x,y) \in \set{A} \right\rbrace.
\end{equation}
Alternatively, for all sets~$\set{B} \in \mathscr{F}_{\set{Y}} \times \mathscr{F}_{\set{X}}$, it follows that
\begin{equation}
\label{EqNovember10at20h22in2024Nice}
P_{ YX } \left( \set{B} \right) = \int P_{X | Y = y} \left( \set{B}_{y}\right) \mathrm{d} P_{Y} \left( y\right),
\end{equation}
where~$\set{B}_{y}$ is the section set of~$\set{B}$ determined by $y$.

Within this context, the following folklore theorem highlights a property of conditional measures, which is reminiscent of the unit measure axiom in probability theory. 
\begin{theorem}[Unit Measure]\label{ThUM}
Consider the conditional probability measures~$P_{Y | X} \in \triangle\left( \set{Y} | \set{X} \right)$ and~$P_{X | Y}\in \triangle\left( \set{X} | \set{Y} \right)$; the probability measures~$P_{Y} \in \triangle\left( \set{Y} \right)$ 
and~$P_{X} \in \triangle\left(  \set{X} \right)$ that satisfy~\eqref{eqmarginal1} and~\eqref{eqmarginal2}. Assume that for all~$x \in \set{X}$, the probability measure~$\abscont{P_{Y | X = x}}{P_{Y}}$. Then,
\begin{IEEEeqnarray}{rCl}
\label{EqThCondPorp}
\int  \frac{\mathrm{d} P_{Y|X = x}}{\mathrm{d}P_{Y}} (y) \mathrm{d}P_{X}(x) &\eqas{P_Y}& 1.
 \end{IEEEeqnarray}
 \end{theorem}
\begin{IEEEproof}
For all $\set{A} \in \mathscr{F}_{\set{Y}}$, from \eqref{eqmarginal1}, it holds that
\begin{IEEEeqnarray}{rCl}
\label{EqDefOfMarg}
 P_{Y}(\set{A}) & =& \iint_{\set{A}} \mathrm{d}P_{Y | X = x}(y) \mathrm{d}P_X (x)\\
 \label{EqIDK}
  & = & \iint_{\set{A}} \frac{\mathrm{d} P_{Y|X = x}}{\mathrm{d}P_{Y}}(y) \mathrm{d}P_{Y}(y) \mathrm{d}P_X (x)\\
  \label{EqResult}
  & =& \int_{\set{A}} \int  \frac{\mathrm{d} P_{Y|X = x}}{\mathrm{d}P_{Y}}(y) \mathrm{d}P_X (x) \mathrm{d}P_{Y}(y),
 \end{IEEEeqnarray}
where the equality in  \eqref{EqIDK} follows from a change of measure (Theorem~\ref{ThCOM}). Moreover, \eqref{EqResult} is 
obtained using Fubini's theorem \cite[Theorem~2.6.6]{ash2000probability}.
The proof proceeds by noticing that $P_{Y}(\set{A})  =\int _{\set{A}} \mathrm{d}P_{Y}(y)$, and thus from Theorem~\ref{ThRNT} and the equality in \eqref{EqResult}, the statement in \eqref{EqThCondPorp} holds.
\end{IEEEproof}

 The following folklore theorems are reminiscent of the Bayes rule.

\begin{theorem}[Bayes-like rule]\label{TheoBYR}
Consider the conditional probability measures~$P_{Y | X}$ and~$P_{X | Y}$; the probability measures~$P_{Y}$ 
and~$P_{X}$ that satisfy~\eqref{eqmarginal1} and~\eqref{eqmarginal2}; and the joint probability measures~$P_{YX}$ and~$P_{XY}$ in \eqref{EqOctober30at7h48in2024SophiaAntipolis} and \eqref{EqJun3at14h31in2024} respectively. 
Let also $P_{X}P_{Y} \in \triangle\left( \set{X} \times \set{Y} , \mathscr{F}_{\set{X}} \times \mathscr{F}_{\set{Y}}\right)$ and $P_{Y}P_{X} \in \triangle\left( \set{Y} \times \set{X} , \mathscr{F}_{\set{Y}} \times \mathscr{F}_{\set{X}}\right)$ be the measures formed by the product of the marginals $P_{X}$ and $P_{Y}$. 
 Assume that: $\left(a\right)$ For all~$x \in \set{X}$,~$\abscont{P_{Y | X = x}}{P_{Y}}$; 
and $\left(b\right)$ For all~$y \in \set{Y}$,~$\abscont{P_{X | Y = y}}{P_{X}}$.
Then,
\begin{IEEEeqnarray}{rCl}
\label{EqOuter1Inner1}
\frac{\mathrm{d} P_{XY}}{\mathrm{d} P_{X}P_{Y}} \left( x,y \right) \eqas{P_XP_Y} \frac{\mathrm{d} P_{X | Y = y}}{\mathrm{d} P_{X}} \left( x \right)
  \eqas{P_XP_Y} \frac{\mathrm{d} P_{Y | X = x}}{\mathrm{d} P_{Y}}\left( y \right)
 \label{EqInner2Outer2}
  \eqas{P_XP_Y}  \frac{\mathrm{d} P_{YX}}{\mathrm{d} P_{Y}P_{X}}\left( y,x\right).
\end{IEEEeqnarray}
\end{theorem}
\begin{IEEEproof}
Note that assumptions $\left(a\right)$ 
and $\left(b\right)$ are sufficient for the Radon-Nikodym derivatives of $P_{XY}$ with respect to $P_XP_Y$; and~$P_{YX}$ with respect to $P_YP_X$ to exist. Hence, it follows that for all sets~$\set{A}~\in~ \mathscr{F}_{\set{X}} \times \mathscr{F}_{\set{Y}}$,
\begin{IEEEeqnarray}{rCl}
\label{EqJanuary19at21h57inDubai}
P_{XY }\left( \set{A} \right) & = & \int_{\set{A}} \frac{\mathrm{d}P_{XY}} {\mathrm{d}P_{X}P_{Y}} (x,y) \mathrm{d}P_{X}P_{Y} (x,y),  \IEEEeqnarraynumspace 
\end{IEEEeqnarray}
which follows from Theorem~\ref{ThCOM}.
Note also that from~\eqref{EqJun3at14h31in2024}, it follows that
\begin{IEEEeqnarray}{rCl}
\label{EqOctober31at13h12in2024Nicea}
P_{X Y }\left( \set{A} \right) & = & \iint_{ \set{A}_{y} } \mathrm{d} P_{X | Y = y} \left( x \right) \mathrm{d} P_{Y} \left( y\right)\\
\label{EqOctober31at13h12in2024Niceb}
&=& \iint_{ \set{A}_{y} } \frac{\mathrm{d} P_{X | Y = y} }{\mathrm{d}P_{X}} \left( x \right) \mathrm{d}P_{X}\left( x \right) 
\mathrm{d} P_{Y} \left( y\right) \IEEEeqnarraynumspace \\
\label{EqNovember15at16h25in2024InTheBusToNiceA}
& = & \iint \mathds{1}_{\set{A}_y}(x) \frac{\mathrm{d} P_{X | Y = y} }{\mathrm{d}P_{X}} \left( x \right)  \mathrm{d}P_{X}\left( x \right) 
\mathrm{d} P_{Y} \left( y\right) \IEEEeqnarraynumspace \\
\label{EqNovember15at16h25in2024InTheBusToNiceB}
& = & \int \mathds{1}_{\set{A}}(x,y) \frac{\mathrm{d} P_{X | Y = y} }{\mathrm{d}P_{X}} \left( x \right)  \mathrm{d} P_{X} P_{Y} \left(x, 
y\right) \IEEEeqnarraynumspace \\
\label{EqOctober31at13h12in2024Nicec}
& = & \int_{\set{A}} \frac{\mathrm{d} P_{X | Y = y} }{\mathrm{d}P_{X}} \left( x \right) \mathrm{d} P_{X} P_{Y} \left(x,  y\right), 
\IEEEeqnarraynumspace
\end{IEEEeqnarray}
where, the set~$\set{A}_{y}$ is defined in~\eqref{EqNovember15at16h39in2024InTheBusToNice}.  Moreover, the equality 
in~\eqref{EqOctober31at13h12in2024Niceb} follows from Assumption~$\left(b\right)$ and Theorem \ref{ThRNT}. 
Similarly, from~\eqref{EqMay20at14h23in2024}, it follows that 
\begin{IEEEeqnarray}{rCl}
\label{EqNovember15at16h36in2024InTheBusToNiceA}
P_{X Y }\left( \set{A} \right) & = & \iint_{ \set{A}_{x} } \mathrm{d} P_{Y | X = x} \left( y \right) \mathrm{d} 
P_{X} \left( x\right)\\
\label{EqNovember15at16h36in2024InTheBusToNiceB}
& = & \iint_{ \set{A}_{x} } \frac{\mathrm{d} P_{Y | X = x} }{\mathrm{d}P_{Y}} \left( y \right)\mathrm{d} P_{Y} \left( y\right) \mathrm{d}P_{X}\left( x \right)  \IEEEeqnarraynumspace \\
\label{EqNovember15at16h36in2024InTheBusToNiceC}
& = & \iint \mathds{1}_{\set{A}_x}(y) \frac{\mathrm{d} P_{Y | X = x} }{\mathrm{d}P_{Y}} \left( y \right) \mathrm{d} P_{Y} \left( y\right) 
\mathrm{d}P_{X}\left( x \right) \IEEEeqnarraynumspace \\
\label{EqNovember15at16h36in2024InTheBusToNiceC1}
& = & \iint \mathds{1}_{\set{A}_y}(x) \frac{\mathrm{d} P_{Y | X = x} }{\mathrm{d}P_{Y}} \left( y \right)  \mathrm{d}P_{X}\left( x \right) 
\mathrm{d} P_{Y} \left( y\right)  \IEEEeqnarraynumspace \\
\label{EqNovember15at16h36in2024InTheBusToNiceD}
& = & \int \mathds{1}_{\set{A}}(x,y) \frac{\mathrm{d} P_{Y | X = x} }{\mathrm{d}P_{Y}} \left( y \right) \mathrm{d} P_{X}P_{Y}  \left(  x , 
y\right)  \IEEEeqnarraynumspace \\
\label{EqNovember15at16h36in2024InTheBusToNiceE}
& = & \int_{\set{A}} \frac{\mathrm{d} P_{Y | X = x} }{\mathrm{d}P_{Y}} \left( y \right)  \mathrm{d} P_{X}P_{Y}  \left(  x , y\right),  
\IEEEeqnarraynumspace
\end{IEEEeqnarray}
where, the set~$\set{A}_{x}$ is defined in~\eqref{EqNovember15at16h57in2024InTheBusToNice}.  Moreover, the equality 
in~\eqref{EqNovember15at16h36in2024InTheBusToNiceB} follows from Assumption~$\left(a\right)$ and Theorem \ref{ThRNT}; and
the equality in~\eqref{EqNovember15at16h36in2024InTheBusToNiceC1} follows by exchanging the order of integration \cite[Theorem~2.6.6]{ash2000probability}.
Finally, from~\eqref{EqJanuary10at16h43in2025}, it follows that
\begin{IEEEeqnarray}{rCl}
\nonumber
P_{XY}\left( \set{A} \right) 
&= &\int_ {\hat{\set{A}} } \mathrm{d} P_{YX}\left(y,x\right)\\
\label{EqAntibesTT1}
&=& \int_{\hat{\set{A}}}\frac{ \mathrm{d}P_{YX}} { \mathrm{d} P_YP_X}\left(y,x\right) \mathrm{d}P_YP_X\left(y,x\right)\\
&=& \int \mathds{1}_{\hat{\set{A}}}(y,x) \frac{ \mathrm{d}P_{YX}} { \mathrm{d} P_YP_X}\left(y,x\right) \mathrm{d}P_YP_X\left(y,x\right) \IEEEeqnarraynumspace \\
\label{EqAntibesTT2}
&=&  \iint \mathds{1}_{\hat{\set{A}}_x}(y)\frac{ \mathrm{d}P_{YX}} { \mathrm{d} P_YP_X}\left(y,x\right) \mathrm{d}P_Y(y) \mathrm{d}P_X(x)\IEEEeqnarraynumspace \\
\label{EqAntibesTT3}
&=& \iint \mathds{1}_{\hat{\set{A}}_y}(x)\frac{ \mathrm{d}P_{YX}} { \mathrm{d} P_YP_X}\left(y,x\right) \mathrm{d}P_X(x) \mathrm{d}P_Y(y) \IEEEeqnarraynumspace \\
&=& \int \mathds{1}_{\set{A}}(x,y)\frac{ \mathrm{d}P_{YX}} { \mathrm{d} P_YP_X}\left(y,x\right) \mathrm{d}P_XP_Y(x,y)\\ \IEEEeqnarraynumspace
\label{EqAntibesin2025Jan}
&=& \int_{\set{A}} \frac{ \mathrm{d}P_{YX}} { \mathrm{d} P_YP_X}\left(y,x\right) \mathrm{d}P_XP_Y(x,y), \IEEEeqnarraynumspace
\end{IEEEeqnarray}
where the set $\hat{\set{A}}$ is defined in \eqref{EqNovember16at18h49in2024Nice}. The equality in \eqref{EqAntibesTT1} is obtained by performing a change of measure using Theorem~\ref{ThCOM} under assumption $\left(a\right)$; and the equality in \eqref{EqAntibesTT3} follows by exchanging the order of integration \cite[Theorem~2.6.6]{ash2000probability}. 

The proof is completed from Theorem~\ref{ThRNT} and by combining equations~\eqref{EqJanuary19at21h57inDubai}, ~\eqref{EqOctober31at13h12in2024Nicec}, ~\eqref{EqNovember15at16h36in2024InTheBusToNiceE} and~\eqref{EqAntibesin2025Jan}, which establish \eqref{EqOuter1Inner1}.
\end{IEEEproof}

\begin{theorem}[Inverse Bayes-like Rule]\label{TheoBYR2}
Consider the conditional probability measures~$P_{Y | X}$ and~$P_{X | Y}$;  and the probability measures~$P_{Y}$ 
and~$P_{X}$ that satisfy~\eqref{eqmarginal1} and~\eqref{eqmarginal2}; and the joint probability measures~$P_{YX}$ and~$P_{XY}$ in~\eqref{EqOctober30at7h48in2024SophiaAntipolis} and~\eqref{EqJun3at14h31in2024}  respectively.
 Assume that: $\left(a\right)$ For all~$x \in \set{X}$,~$\abscont{P_{Y}}{P_{Y | X = x}}$; 
and $\left(b\right)$ For all~$y \in \set{Y}$,~$\abscont{P_{X}}{P_{X | Y = y}}$.  
Then, 
\begin{IEEEeqnarray}{rCl}
 \label{EqInverseInner1Inner2}
  \frac{\mathrm{d} P_{X}P_{Y}}{\mathrm{d} P_{XY}} \left( x,y \right) \eqas{P_{XY}} \frac{\mathrm{d} P_{X}}{\mathrm{d} P_{X | Y = y}} \left( x \right)   \eqas{P_{XY}} 
 \frac{\mathrm{d} P_{Y}}{\mathrm{d} P_{Y | X = x}}\left( y \right) \eqas{P_{YX}}  \frac{\mathrm{d} P_{Y}P_{X}}{\mathrm{d} P_{YX}}\left( y,x\right).
\end{IEEEeqnarray}
\end{theorem}

\begin{IEEEproof}
The proof follows along the same lines as the proof of Theorem~\ref{TheoBYR}.

Note that assumptions $\left(a\right)$ 
and $\left(b\right)$ are sufficient for the Radon-Nikodym derivatives of $P_XP_Y$ with respect to $P_{XY}$; and~$P_YP_X$ with respect to $P_{YX}$ to exist. Hence, it follows that for all sets~$\set{A}~\in~ \mathscr{F}_{\set{X}} \times \mathscr{F}_{\set{Y}}$,
\begin{IEEEeqnarray}{rCl}
\label{EqJuly7inSaintFlourat16h57}
P_{X}P_{Y} \left( \set{A} \right) & = & \int_{\set{A}}  \RND{P_{X}P_{Y}}{P_{XY}} (x,y) \mathrm{d}P_{XY} (x,y),
\end{IEEEeqnarray}
which follows from Theorem~\ref{ThCOM}.

The proof follows by observing that for all measurable sets~$\set{A} \in \set{X} \times \set{Y}$,  the product measure~$P_{X}P_{Y} \in \triangle\left( \set{X} \times \set{Y} \right)$ satisfies
\begin{eqnarray}
\nonumber
P_{X}P_{Y}  \left(\set{A} \right)  
\label{EqNovember16at18h29in2024NiceA}
& = &\int_{\set{A}} \mathrm{d}P_{X}P_{Y}  \left( x, y\right) \\
\label{EqNovember16at18h29in2024NiceB}
& = &\int \int_{\set{A}_{y}} \mathrm{d}P_{X}\left( x\right)  \mathrm{d}P_{Y}  \left(y\right) \\
\label{EqNovember16at18h29in2024NiceC}
& = &\int \int_{\set{A}_{y}}  \frac{\mathrm{d}P_{X}}{\mathrm{d}P_{X | Y = y}} \left( x \right) \mathrm{d}P_{X | Y = y} \left( x \right)  \mathrm{d}P_{Y}  \left(y\right) \spnum \\
\label{EqNovember16at18h29in2024NiceD}
& = &\iint  \mathds{1}_{\set{A}_y}(x) \frac{\mathrm{d}P_{X}}{\mathrm{d}P_{X | Y = y}} \left( x \right)  \mathrm{d}P_{X | Y = y} \left( x \right)  \mathrm{d}P_{Y}  \left(y\right) \spnum \\
\label{EqNovember16at18h29in2024NiceE}
& = &\int \mathds{1}_{\set{A}_y}(x) \frac{\mathrm{d}P_{X}}{\mathrm{d}P_{X | Y = y}} \left( x \right) \mathrm{d}P_{XY} \left( x, y\right) \spnum \\
\label{EqNovember16at18h29in2024NiceF}
& = &\int_{\set{A}} \frac{\mathrm{d}P_{X}}{\mathrm{d}P_{X | Y = y}} \left( x \right)    \mathrm{d}P_{XY} \left( x, y\right), \spnum 
\end{eqnarray}
where the set~$\set{A}_{y}$ is defined in~\eqref{EqNovember15at16h39in2024InTheBusToNice}; the equality in~\eqref{EqNovember16at18h29in2024NiceC} follows from Assumption~$(b)$ and \cite[Theorem~2.2.3]{lehmann2005testing}; and the measure~$P_{XY}$ is defined in~\eqref{EqJun3at14h31in2024}.

The proof proceeds by noticing that for all measurable sets~$\set{A} \in \set{X} \times \set{Y}$,  the product measure~$P_{X}P_{Y} \in \triangle\left( \set{X} \times \set{Y} \right)$ also satisfies
\begin{eqnarray}
\nonumber
P_{X}P_{Y}  \left(\set{A} \right)  
\label{EqNovember16at18h36in2024NiceA}
& = &\int_{\set{A}} \mathrm{d}P_{X}P_{Y}  \left( x, y\right) \\
\label{EqNovember16at18h36in2024NiceB}
& = &\int \int_{\set{A}_{y}} \mathrm{d}P_{X}\left( x\right)  \mathrm{d}P_{Y}  \left(y\right) \\
\label{EqNovember16at18h36in2024NiceC}
& = &\int \int_{\set{A}_{x}}  \mathrm{d}P_{Y}  \left(y\right) \mathrm{d}P_{X}\left( x\right)  \\
\label{EqNovember16at18h36in2024NiceC1}
& = &\int \int_{\set{A}_{x}}  \frac{\mathrm{d}P_{Y}}{\mathrm{d}P_{Y | X = x}} \left( y \right) \mathrm{d}P_{Y | X = x} \left( y \right)  \mathrm{d}P_{X}  \left(x\right) \spnum \\
\label{EqNovember16at18h36in2024NiceD}
& = &\iint \mathds{1}_{\set{A}_x}(y) \frac{\mathrm{d}P_{Y}}{\mathrm{d}P_{Y | X = x}} \left( y \right) \mathrm{d}P_{Y | X = x} \left( y \right)  \mathrm{d}P_{X}  \left(x\right) \spnum \\
\label{EqNovember16at18h36in2024NiceE}
& = &\int  \mathds{1}_{\set{A}_x}(y) \frac{\mathrm{d}P_{Y}}{\mathrm{d}P_{Y | X = x}} \left( y \right) \mathrm{d}P_{YX}  \left(y, x\right) \spnum \\
\label{EqNovember16at18h36in2024NiceF}
& = &\int_{\hat{\set{A}}}   \frac{\mathrm{d}P_{Y}}{\mathrm{d}P_{Y | X = x}} \left( y \right) \mathrm{d}P_{YX}  \left(y, x\right) \spnum \\
\label{EqNovember16at18h36in2024NiceG}
& = &\int_{\set{A}}   \frac{\mathrm{d}P_{Y}}{\mathrm{d}P_{Y | X = x}} \left( y \right) \mathrm{d}P_{XY}  \left(x,y\right), \spnum 
\end{eqnarray}
where the sets~$\set{A}_{x}$,~$\set{A}_{y}$, and~$\hat{\set{A}}$ are defined in~\eqref{EqNovember15at16h57in2024InTheBusToNice},~\eqref{EqNovember15at16h39in2024InTheBusToNice}, and~\eqref{EqNovember16at18h49in2024Nice}; and the measure~$P_{YX}$ is defined in~\eqref{EqOctober30at7h48in2024SophiaAntipolis}.
The equality in~\eqref{EqNovember16at18h36in2024NiceC} follows by exchanging the order of the integrals \cite[Theorem~2.6.6]{ash2000probability}; 
the equality in~\eqref{EqNovember16at18h36in2024NiceC1} follows from Assumption~$(a)$ and \cite[Theorem~2.2.3]{lehmann2005testing}.

Finally, for all measurable sets~$\set{A} \in \set{X} \times \set{Y}$ and $\hat{\set{A}}$ in~\eqref{EqNovember16at18h49in2024Nice},
\begin{IEEEeqnarray}{rCl}
P_{X}P_{Y} \left( \set{A} \right) 
&=& \int_{\hat{\set{A}}}   \mathrm{d}P_{Y}P_{X}(y,x)\\
&=& \int_{\hat{\set{A}}}  \RND{P_{Y}P_{X}}{P_{YX}} (y,x) \mathrm{d}P_{YX}(y,x)\\
\label{EqJuly7inSaintFlourat16h58}
&=& \int_{\set{A}}  \RND{P_{Y}P_{X}}{P_{YX}} (y,x) \mathrm{d}P_{XY}(x,y).
\end{IEEEeqnarray}

The proof is completed from Theorem~\ref{ThRNT} and by combining equations~\eqref{EqJuly7inSaintFlourat16h57}, ~\eqref{EqNovember16at18h29in2024NiceF}, ~\eqref{EqNovember16at18h36in2024NiceG} and~\eqref{EqJuly7inSaintFlourat16h58}, which establish \eqref{EqInverseInner1Inner2}.
\end{IEEEproof}

\begin{remark}
An alternative proof for Theorem \ref {ThUM} can be written as follows, by combining Theorems~\ref{TheoInverseRND} and~\ref{TheoBYR2}. 
\begin{eqnarray}
\label{EqSaintFlourJuly2ndat19h15}
 \int  \frac{\mathrm{d} P_{Y|X = x}}{\mathrm{d}P_{Y}} (y) \mathrm{d}P_{X}(x)
 &=& \int  \frac{\mathrm{d} P_{Y|X = x}}{\mathrm{d}P_{Y}} (y)\frac{\mathrm{d} P_{X}}{\mathrm{d}P_{X | Y= y}} (x) \mathrm{d}P_{X|Y = y}(x)\\\label{EqSaintFlourJuly2ndat19h17}
 &=& \int  \frac{\mathrm{d} P_{Y|X = x}}{\mathrm{d}P_{Y}} (y)\frac{\mathrm{d} P_{Y}}{\mathrm{d}P_{Y | X= x}} (y) \mathrm{d}P_{X|Y = y}(x)\\
 \label{EqSaintFlourJuly2ndat19h18}
 &=&  \int  \mathrm{d}P_{X|Y = y}(x) = 1, 
 \end{eqnarray}
 where the equality in \eqref{EqSaintFlourJuly2ndat19h15} follows from Theorem~\ref{ThCOM}; the equality in \eqref{EqSaintFlourJuly2ndat19h17} follows from Theorem~\ref{TheoBYR2} and the equality in \eqref{EqSaintFlourJuly2ndat19h18} follows form Theorem~\ref{TheoInverseRND}.
\end{remark}

The following theorem introduces an identity involving the sum of a mutual information \cite{Shannon-1948a, Shannon-1948b} and a lautum information \cite{palomar2008lautum}. Such a result appeared first in \cite[Corollary~13]{perlaza2024generalization} as a byproduct of the properties of the generalization error of machine learning algorithms. In this work, a novel proof is presented in a general context exclusively using the folklore theorems on the RND described above.

\begin{theorem}
\label{TheoWhatever}
Consider the conditional probability measure $P_{Y | X}$;   the probability measures~$P_{Y}$ 
and~$P_{X}$ that satisfy~\eqref{eqmarginal1} and~\eqref{eqmarginal2}; and the joint probability measure~$P_{YX}$ in~\eqref{EqOctober30at7h48in2024SophiaAntipolis}. Let $Q$ be some $\sigma$-finite measure in $\simplex{\set{Y}}$ and 
 assume that for all~$x \in \set{X}$,~$P_Y\ll P_{Y|X=x}\ll Q \ll P_{Y}$. 
Then, 
\begin{IEEEeqnarray}{rCl}
\nonumber
\mathrm{I}\autoparent{ P_{Y|X} ; P_X} + \mathrm{L}\autoparent{ P_{Y|X} ; P_X}  
&=& \iint \log \autoparent{\RND{P_{Y|X=x} }{Q}(y)} \mathrm{d}P_{Y|X=x} (y) \mathrm{d}P_X(x)\\
\label{EqThmWhatever}
&-&  \iint \log \autoparent{\RND{P_{Y|X=x} }{Q}(y)} \mathrm{d}P_{Y}(y) \mathrm{d}P_X(x),
\end{IEEEeqnarray}
where $\mathrm{I}\autoparent{ P_{Y|X} ; P_X}$ and $\mathrm{L}\autoparent{ P_{Y|X} ; P_X}$ are respectively the mutual information and lautum information induced by the joint probability measure~$P_{YX}$ in~\eqref{EqOctober30at7h48in2024SophiaAntipolis}.
\end{theorem}
\begin{IEEEproof}
From the definition of mutual information, it follows that
\begin{IEEEeqnarray}{rCl}
\label{EqProofThmWhateverMutual1}
&&\mathrm{I}\autoparent{ P_{Y|X} ; P_X} = \iint \log \autoparent{\RND{P_{Y|X=x} }{P_Y}(y)} \mathrm{d}P_{Y|X=x} (y) \mathrm{d}P_X(x) \middlesqueezeequ \spnum \\
\label{EqProofThmWhateverMutual2}
&=& \iint \log \autoparent{\RND{P_{Y|X=x} }{Q}(y) \RND{Q}{P_Y}(y)} \mathrm{d}P_{Y|X=x} (y) \mathrm{d}P_X(x) \\
\nonumber 
&=& \iint \log \autoparent{\RND{P_{Y|X=x} }{Q}(y) } \mathrm{d}P_{Y|X=x} (y) \mathrm{d}P_X(x)\\
\label{EqProofThmWhateverMutual3}
&+& \iint \log \autoparent{\RND{Q}{P_Y}(y)} \mathrm{d}P_{Y|X=x} (y) \mathrm{d}P_X(x),
\end{IEEEeqnarray}
where the equality in \eqref{EqProofThmWhateverMutual2} follows from Theorem~\ref{TheoChainsAndBlood}. The proof follows by re-writing the second term in~\eqref{EqProofThmWhateverMutual3}. That is,
\begin{IEEEeqnarray}{rCl}
\nonumber
\label{EqProofThmWhateverMutualInner1}
&& \iint \log \autoparent{\RND{Q}{P_Y}(y)} \mathrm{d}P_{Y|X=x} (y) \mathrm{d}P_X(x) \\
\label{EqProofThmWhateverMutualInner2}
&=& \iint \autoparent{ \log \autoparent{\RND{Q}{P_Y}(y)} \RND{P_{Y|X=x}}{P_Y}(y)} \mathrm{d}P_Y(y) \mathrm{d}P_X(x)\\
\label{EqProofThmWhateverMutualInner3}
&=& \int \log \autoparent{\RND{Q}{P_Y}(y)} \autoparent{\int \RND{P_{Y|X=x}}{P_Y}(y) \mathrm{d}P_X(x)}\mathrm{d}P_Y(y) \spnum\\
\label{EqProofThmWhateverMutualInner4}
 &=& \int \log \autoparent{\RND{Q}{P_Y}(y)} \mathrm{d}P_Y(y),
\end{IEEEeqnarray}
where the equality in \eqref{EqProofThmWhateverMutualInner2} follows from Theorem~\ref{ThCOM}; the equality in \eqref{EqProofThmWhateverMutualInner3} is a reorganization of terms in the integrals; and the equality in \eqref{EqProofThmWhateverMutualInner4} follows from Theorem~\ref{ThUM}.
From the definition of lautum information, it follows that
\begin{IEEEeqnarray}{rCl}
\nonumber
 \mathrm{L}\autoparent{ P_{Y|X} ; P_X} & = &  \iint \log \autoparent{\RND{P_Y}{P_{Y|X=x} }(y)} \mathrm{d}P_{Y}(y) \mathrm{d}P_X(x)\\
\label{EqProofThmWhateverLautum2}
&=& \iint \log \autoparent{\RND{P_Y}{Q }(y) \RND{Q}{P_{Y|X=x} }(y)} \mathrm{d}P_{Y}(y) \mathrm{d}P_X(x)\\
\nonumber
&=&  \int \log \autoparent{\RND{P_Y}{Q }(y)}\mathrm{d}P_{Y}(y)\\
\label{EqProofThmWhateverLautum3}
&& -  \iint \log \autoparent{\RND{P_{Y|X=x} }{Q}(y)} \mathrm{d}P_{Y}(y) \mathrm{d}P_X(x), \spnum
\end{IEEEeqnarray}
where \eqref{EqProofThmWhateverLautum2} follows from Theorem~\ref{TheoChainsAndBlood} and \eqref{EqProofThmWhateverLautum3} is a direct consequence of the properties of the logarithmic function and Theorem~\ref{TheoInverseRND}.
The proof is completed by plugging \eqref{EqProofThmWhateverMutualInner4} in \eqref{EqProofThmWhateverMutual3}; and adding the resulting \eqref{EqProofThmWhateverMutual3} to \eqref{EqProofThmWhateverLautum3}.  
\end{IEEEproof}
The relevance of expressing the sum of a mutual information and a lautum information in terms of a free parameter (i.e., the measure $Q$ in \eqref{EqThmWhatever}) lies in a hypothesis-testing interpretation for the term 
 $\log \autoparent{\RND{P_{Y| X = x}}{Q}(y)}$, which can be viewed as a log-likelihood ratio for distinguishing the hypothesis $Y \sim P_{Y|X = x}$ from $Y \sim Q$, given some data point $(x,y)$. Consequently, such a sum captures the variation of the expectation of such a log-likelihood ratio when the ground truth probability measure shifts  from $P_{Y}P_{X}$ to $P_{XY}$.
This interpretation has played a central role for establishing bridges between the generalization  theory and hypothesis testing~\cite{perlaza2024generalization}. 

\phantomsection                      
\addcontentsline{toc}{section}{References}
\bibliographystyle{IEEEtranlink}
\bibliography{../../../../Latexfiles/Bibliography/BermudezRR.bib}
\end{document}